\documentstyle[osa,manuscript]{revtex}

\begin{document}

\begin{center}{\Large \bf Necessary and sufficient condition for hydrostatic equilibrium in general relativity} \\
\vskip.25in
{P. S. Negi}

{\it
Department of Physics, Kumaun University,
Nainital 263 002, India} %
\end{center}


\begin{abstract}                
We present explicit examples to show that the 
   `compatibility criterion' [recently obtained by us towards
providing equilibrium configurations compatible with the structure of general relativity]
which states that: for a given value of $\sigma [\equiv (P_0/E_0) \equiv $ the ratio  of  central  pressure  to   central   energy-density],   
the compactness ratio $u [\equiv (M/R)$, where $M$ is the total mass and $R$ is the radius of the configuration]   of  any 
static configuration cannot  exceed the compactness ratio, $u_h$, of the homogeneous density sphere (that is, $u \leq  u_h$), is capable of providing a {\em necessary} and {\em sufficient} condition for any
regular configuration to be compatible with the state of hydrostatic equilibrium. This conclusion
is drawn on the basis of the finding that the $M-R$ relation gives the necessary and sufficient condition
for dynamical stability of equilibrium configurations only when the compatibility criterion for
these configurations is appropriately satisfied. In this regard, we construct an appropriate sequence composed of core-envelope models
on the basis of compatibility criterion, such that each member of this sequence satisfies the extreme case of causality condition $v = c = 1$ at the centre. 
The maximum stable value of $u \simeq 0.3389$ (which occurs for the model corresponding to the maximum value of mass in
the mass-radius relation) and the corresponding central value of the local adiabatic index, $(\Gamma_1)_0 \simeq 2.5911$,
of this model are found fully consistent with those of the corresponding
{\em absolute} values, $u_{\rm max} \leq 0.3406$, and $(\Gamma_1)_0 \leq 2.5946$, which impose strong constraints on
these parameters of such models. In addition to this example, we also study dynamical stability of pure adiabatic polytropic configurations
on the basis of variational method for the choice of the `trial function', $\xi =re^{\nu/4}$, as well as the mass-central density
relation, since the compatibility criterion is appropriately satisfied for these models. The results of
this example provide additional proof in favour of the statement regarding compatibility criterion mentioned above.
Together with other results, this study also confirms the previous claim that just the choice of the `trial function', $\xi =re^{\nu/4}$,
 is capable of providing the necessary and sufficient condition
for dynamical stability of a mass on the basis of variational method. Obviously, the upper bound on compactness ratio of 
neutron stars, $u \cong 0.3389 $, which belongs to two-density model studied here, turns out to be much
stronger than the corresponding `absolute' upper bound mentioned in the literature.


\end{abstract}
\newpage

\section{ INTRODUCTION}

Einstein's   field    equations for static and spherically symmetric mass distribution 
were first solved  by Schwarzschild (1916).
The first solution describes the geometry of the space-time exterior to a prefect
fluid sphere in hydrostatic equilibrium. While the other, known as interior Schwarzschild solution, corresponds to the interior geometry
of a fluid sphere of constant (homogeneous) energy-density, $E$. The importance of these two
solutions in general relativity is well known. The interior Schwarzschild solution 
provides  two  very  important  features  towards  
obtaining configurations in  hydrostatic  equilibrium,  compatible  
with general relativity, namely - (i) It gives an  absolute  upper  
limit on compactness ratio, $u (\equiv M/R$, mass  to  size  ratio  of  
the entire configuration in geometrized units) $\leq (4/9)$ for any static and spherical configuration
(belonging to arbitrary density profiles, provided the density does not increase outwards) in hydrostatic equilibrium  
(Buchdahl 1959; Weinberg 1972), and (ii) For an assigned value of the  compactness  ratio,  
$u$, and radius $R$ (or mass $M$),  the  minimum  central  pressure, $ P_0$,   corresponds   to   the  
homogeneous density solution (see,  e.g., Weinberg 1972). 
 
      Despite the non linear and coupled differential equations, various  exact  
solutions of the field equations for  static  and  spherically  symmetric   metric   are  
available  in  the  literature (see, e. g., Kramer et al 1980) 
 which may be 
used to obtain  various  physical  properties  of  spherical  and 
static compact object (provided they are physically realistic).  Knutsen (1988; 1989)  
examined  physical  
properties of the various exact solutions and found that  these  solutions  
correspond to nice physical  properties  and  also  remain  stable  
against small radial pulsations upto certain values of $u$.

Another way to explore the  physical  properties  of compact objects
like neutron star, one may expect to have some  physically
viable equation of state (EOS). However,  for  such  objects  the 
equations of state (EOSs) are not well known  [empirically]  because  of 
the lack of knowledge of nuclear interactions beyond the  density 
$ \sim 10^{14} {\rm g\, cm}^{-3}$ (Dolan 1992), and the only way to obtain EOSs far
beyond  this density  range  is  extrapolation.  Various   such
extrapolated equations  are available in the literature (Arnett \& Bowers 1977).
As  a  way  out,  
one  can  impose  some  restrictions  upon  the  known   physical 
quantities,  such  that,  the   speed   of   sound   inside   the 
configuration, $v [\equiv \sqrt (dP/dE)]$, does  not  exceed  the  speed  of 
light in vacuum, i.e., $v \leq c =  1$  (in  geometrized  units),  and 
obtain various physical properties, like upper bound on stable neutron  star  masses (Rhoades \& Ruffini 1974;
Brecher \& Caporaso 1976; Hartle 1978; Friedman \& Ipser 1987).
Haensel and Zdunik (1989) have shown that the  only  EOS  which  can 
describe a submillisecond pulsar and the static mass of $1.442  M_\odot $  
simultaneously, corresponds to the EOS, $(dP/dE) = 1$, however, they 
emphasized that this EOS represents an `abnormal' state of matter 
in the sense that pressure vanishes at densities of the order  of 
nuclear density or even higher. 

We have recently proposed a 
core-envelope model with stiffest EOS, $(dP/dE) = 1$, forming the
core and a 
polytropic equation with constant adiabatic index $\Gamma_1$ = (d$lnP/$d$ln \rho)$ [where $P$ is the pressure and 
$\rho$ represents the rest-mass density] describing the envelope, such that $P, E$, both of the metric  parameters ($\nu$, and $\lambda$),
their 
first derivatives, and the speed of sound are  continuous  at  the 
core-envelope boundary  and  at  the  exterior  boundary (surface)  of  the 
structure (Negi \& Durgapal 2000). The other remarkable feature of this core-envelope model is that not only
the `abnormalities' (in the sense discussed in the literature, see, e.g. Lee 1975; Haensel \& Zdunik 1989) disappear, 
the maximum 
value of $u \cong 0.3574$ for the stable configuration turns out to be as large as
that obtained by using the EOS, $(dP/dE) = 1$, alone (see, e. g., Haensel \& Zdunik 1989).
The  maximum  stable mass  of 
neutron star based upon this model (by using the maximum 
value of $u \cong 0.3574$ for stable configuration) turns out to be $7.944 M_\odot$,  if 
the (average) density  of  the  configuration  is  constrained  by 
fastest rotating pulsar, with rotation period, $P_{rot} \cong 1.558$ ms, 
known to date.  The  model  gives  pulsationally  stable 
configurations with compactness ratio $u > (1/3) $, which  are  important  to 
study Ultra-Compact Objects (UCOs) [see, e. g., Negi \& Durgapal 1999a, b; 2000; and references
therein].

Recently, by using property (ii) of the homogeneous density sphere as mentioned
above,  we  have connected  the compactness ratio, $u$, of any static and spherical configuration with the 
corresponding ratio  of  central pressure  to   central   energy-density $\sigma [\equiv (P_0/E_0)$] and worked out  an  
important criterion which concludes that for a given value of $\sigma$, the maximum  
value of compactness ratio, $u(\equiv u_h)$, should always correspond to  the  homogeneous  
density sphere (Negi \& Durgapal 2001).

An examination of this criterion on some well  
known exact solutions and EOSs indicated  that  
this criterion, in fact, is fulfilled only by two types of configurations corresponding to
a {\em single} EOS or density variation: (i) the  {\em regular}
(positive finite density  at  the  origin  which  decreases  
monotonically outwards) configurations
which correspond to a vanishing density at  the  surface  together  
with pressure [so called, the gravitationally-bound structures] (Negi \& Durgapal 2001; Negi 2004b; Negi 2006), and (ii) the structures which correspond
to a non-vanishing surface density but exhibit singularities at the centre, in the sense that both pressure
and density become infinity as $r \rightarrow 0$ [so called, the self-bound singular structures] (Negi 2004b; Negi 2006). 
On the other hand, it is seen that 
the EOSs or analytic  solutions,  corresponding  to  a non-zero 
{\em finite}, surface density (that is,  the  pressure  vanishes  at 
finite surface density, and so called the self-bound regular structures),
in fact, do  not fulfill this criterion (Negi \& Durgapal 2001; Negi 2004b; Negi 2006). We have shown this inconsistency particularly for 
the EOS, $(dP/dE) = 1$ (as it represents the most successful EOS to 
obtain the various extreme characteristics of neutron stars as discussed above).

In addition to the self-bound regular configurations corresponding to
a single density variation, the compatibility criterion may not be satisfied by various two-density, or multiple-density,
regular, gravitationally-bound structures. Such structures are widely discussed in the literature, particularly,
for determining the upper bound on neutron star (NS) masses. In this connection, we would consider the core-envelope
model proposed by Negi \& Durgapal (2000).

The reason(s) behind non-fulfillment of the criterion obtained in the study of Negi and Durgapal (2001)  
by various exact self-bound regular solutions and EOSs, as well as the two-density, gravitationally-bound, regular
structures, and their further implications are discussed in the following sections.

\section{Compatibility criterion: the necessary and sufficient condition for hydrostatic equilibrium
of any static spherical configuration}

In order to bring things together regarding the compatibility of regular
structures mentioned above, we follow
Negi and Durgapal (2001) by assuming a homogeneous   sphere   of   uniform energy-density,
$E$. The equations for  isotropic  pressure  $P$,  and
uniform energy-density $E$, can be  written  in  terms  of  compactness ratio, $u$, and the  radial  coordinate  measured  in  units  of 
configuration size, $y (\equiv r/R)$ as
\begin{eqnarray}
8\pi E R^2    & = & 6u.    \\
8\pi P R^2    & = & 6u \Bigl [\frac{(1-2uy^2)^{1/2} - (1-2u)^{1/2}}{{3(1-2u)}^{1/2} - {(1-2uy^2)}^{1/2}}\Bigr ].
\end{eqnarray}

Let us consider a {\em regular} variable density sphere (with some 
given EOS or analytic solution) with  central  energy-density $E_0$  
and  central  pressure  $P_0$ ,  corresponding  to   the   compactness ratio  $u = u_v$.

Now, we can always construct a  homogeneous  density  sphere 
with  the  same  value  of the compactness ratio $u_v$,   and 
energy-density $E_0$, because if  $P_{0h}$    corresponds  to  the  central 
pressure of this sphere, the ratio $\sigma_h (\equiv P_{0h} /E_0)$ depends only upon 
the assigned value of the compactness ratio $u_v$ .  And,  $P_{0h}$  is 
given by
\begin{equation}
P_{0h} = (6u/8\pi R^2)\Bigl [(1 - (1 - 2u)^{1/2} )/(3(1 - 2u)^{1/2}  - 1) \Bigr ].
\end{equation}
Now, according to property (ii) of homogeneous density sphere, we may write
\begin{equation}
P_0  \geq P_{0h}
\end{equation}
or,
\begin{equation}
(P_0 /E_0) \geq (P_{0h}/E_0).
\end{equation}
Hence for a given value of $u(\equiv u_v)$, we obtain
\begin{equation}
\sigma_v  \geq  \sigma_h
\end{equation}
where $\sigma_v$  is defined as the ratio, $ (P_0 /E_0)$.

Now, varying the compactness ratio, $ u_v $,  for  the  homogeneous 
density sphere from $u_v$  to $u_h$ (say), such that, we should have
\begin{equation}
\sigma_v = \sigma_h.
\end{equation}
For $u = u_h$ , the value of $\sigma_h$  would become
\begin{equation}
\sigma_h  = \Bigl [((1 - 2u_h)^{1/2} - 1)/ (1 - 3(1 - 2u_h)^{1/2} )\Bigr ].
\end{equation}
Substituting Eq. (8)  with  the  help  of  Eq. (7)  into
Eq. (6), we get
\begin{equation}
\Bigl [\frac{(1 - 2u_h)^{1/2} - 1)} {(1 - 3(1 - 2u_h)^{1/2} )}\Bigr ]  \geq \Bigl [\frac {[((1 - 2u_v)^{1/2} - 1)} {(1 - 3(1 - 2u_v)^{1/2} )} \Bigr ].
\end{equation}                                                             
Thus, it is clear from Eq. (9) that
\begin{equation}
u_h  \geq u_v ({\rm \,for \,an \,assigned \,value \,of\,} \sigma).  
\end{equation}
That is, for an assigned value of the ratio of central pressure to 
central  energy-density  $\sigma (\equiv \sigma_v),$  the  compactness ratio   of 
homogeneous density distribution, $u(\equiv u_h)$ should always  be  larger 
than or equal to the compactness ratio $u(\equiv u_v)$ of any  regular
solution\footnote[1]{Notice that this finding is also true for self-bound singular solutions because the ratio of (infinite) central pressure to density turns to be finite [examples of such solutions are well represented by Tolman's type V and VI solutions (Tolman 1939)]. Hence, the notion `any regular' solution may be replaced by `any static' solution [of course, with the requirement that the density decreases monotonically outwards from the centre].}, compatible with the structure of general relativity. Or, 
in other words, for an assigned value of the compactness ratio, 
$u$, the {\em minimum} value of the ratio of central pressure  to  central 
energy-density, $\sigma$, corresponds to the homogeneous density sphere.

In the light of Eq. (10), let us assign the same value $M$ for  the  
total mass corresponding  to  various  regular  configurations  in  
hydrostatic  equilibrium.  If  we  denote  the  density   of   the  
homogeneous sphere by $E_h$ , we can write 
\begin{equation}
E_h   =  3M/(4\pi {R_h}^3)                       
\end{equation}
where $R_h$  denotes the radius of the homogeneous density sphere. 
If $R_v$  represents the radius of any other regular  sphere  for  the  
same mass $M$, the average density $E_v$  of  this  configuration  would  
correspond to 
\begin{equation}
E_v   =  3M/(4\pi {R_v}^3).                                              
\end{equation}
Eq. (10) indicates that $R_v \geq R_h$. By the use of Eqs. (11) and  (12)  
we find that 
\begin{equation}
E_v  \leq E_h.                                                     
\end{equation}
That is, for an assign value of $\sigma$ the  average  energy-density  of  
any regular configuration, $E_v$, should always be less than or equal  
to the density, $E_h$ , of the homogeneous density sphere for the same  
mass $M$. 

 We point out that the regular configurations corresponding to a single exact solution, or EOS
with a  finite  
central and non-vanishing surface density, in fact, do not fulfill the definition of this `actual' total
mass, $M$, which appears in
the  exterior  Schwarzschild  solution [this definition asserts that, being the coordinate mass, the particular 
`type' of density variation considered for it should remain `unknown' to an external observe and this 
is possible only when the mass depends either upon the central density, or upon the surface density, and in any
case, not upon both of them (Negi 2004b; Negi 2006). It follows, therefore, that the central density should be independent of the surface density or vice-versa, according to
the density distribution assigned for the mass]. And the so called self-bound regular structures, 
in fact, violate this requirement as they correspond to a surface density which always depends upon the central
density and vice-versa. Thus, the main findings of the study regarding
this criterion can be summarized in the following manner:

(a) The gravitationally-bound regular configuration and self-bound singular structures, described by 
a single EOS or exact solution,  fulfill the definition of the total mass, $M$, appears
in the exterior Schwarzschild solution, hence the condition of hydrostatic equilibrium is
naturally satisfied by these structures [this finding is fully consistent with the `compatibility
criterion' , because for all (possible) values of $\sigma$, the condition $ u \leq u_h$ is fully satisfied by these configurations].

(b) The self-bound regular configurations, described by a single EOS or exact solution,
can not fulfill the definition of the total mass, $M$, appears in the exterior Schwarzschild
solution, as a result, the state of hydrostatic equilibrium can not be satisfied by them
[this finding is also fully consistent with the `compatibility
criterion' , because such configurations correspond to the condition, $u > u_h$ (for all possible values of $\sigma$)].

(c) The only regular configuration which can exist under the category (b) mentioned above
is described by the homogeneous density distribution.

Note that the two-density or multiple-density models (that is, the structures governed by two or more EOSs assigned
for different regions with appropriate matching conditions at the core-envelope boundaries)
of both of the categories, (a) and (b) described above (such that the definition of the mass $M$ 
mentioned above is appropriately satisfied) are quite possible,
 however, as we will show in the present paper that the fulfillment
of the definition of the mass $`M'$ for any two-density model
represents only a {\em necessary} condition for hydrostatic equilibrium, because the
`compatibility criterion' may not be satisfied by them. 
 As we have noted earlier that the necessary condition for hydrostatic equilibrium  (that is, the fulfillment of the 
definition of the mass $M$) put forward by the exterior
Schwarzschild solution is also sufficient for a {\em single} EOS or exact solution assigned for the mass,
because this fact is also supported by the `compatibility criterion'.
It follows therefore that the `compatibility criterion' is capable of ensuring a sufficient and necessary condition 
for any structure (including two-density or multiple-density distribution)
in the state of hydrostatic equilibrium.

To elaborate this statement more clearly, let us consider the core-envelope model
discussed by Negi and Durgapal (2000). The core of this model is described by an EOS which belongs 
to the category `(b)' mentioned above, and the  matching of various parameters
at the core-envelope boundary is assured by characterizing an envelope which belongs to the category `(a)' EOS. 
That is, `overall' the model describes a gravitationally bound two-density structure [of category `a' mentioned
 above], such that the {\em necessary} condition  for hydrostatic equilibrium
put forward by exterior Schwarzschild solution at the surface of the configuration (that is, the mass, $M$,
depends only upon the central density, meaning thereby that the definition of mass is appropriately satisfied),
even then, the compatibility criterion for hydrostatic equilibrium 
(Negi \& Durgapal 2001) turns out to be unsatisfied for this model (as shown under section 4 of the present study). 
Thus, it follows that the fulfillment of {\em necessary} condition for hydrostatic equilibrium
at the surface, and the achievement of proper matching conditions at the core-envelope
boundary are not {\em sufficient} to assure the condition of hydrostatic equilibrium for any {\em two-density} structure. 
However, the
fulfillment of compatibility criterion alone could provides a {\em necessary} and {\em sufficient} condition for any regular 
configuration
(including two-density structures) to be consistent with the state of hydrostatic equilibrium. 

In order to verify this statement, we would re-investigate the core-envelope model put
forward by Negi and Durgapal (2000), based upon the said compatibility criterion
for hydrostatic equilibrium, such that for each (possible) assigned value of $\sigma$, 
the compactness ratio of the whole configuration, $u$, remains less than or equal to the 
compactness ratio, $u_h$, of the corresponding sphere of homogeneous density distribution.
Such an investigation is possible, because we can re-adjust the boundary, $r_b$, of the 
core-envelope model in such a manner that for an assigned value of $\sigma$, the 
`average density', $E_{av}$(say), of the whole configuration always remains less than
or equal to the density, $E_h$, of the homogeneous density sphere for same mass $M$.
Thus, this criterion should be fulfilled by any regular configuration specified by a single density distribution, a core-envelope model, 
a core-mantle-envelope model, or
any other complicated distribution of matter composed of various regions inside the configuration, in
order to fulfill the state of hydrostatic equilibrium. This statement is verified on the basis of
dynamical stability of some regular configurations, consistent with the compatibility criterion,
in the following sections, and the results which are summarized as Theorem 2 and its subsequent
corollaries in the following section, may be stated in the general form  as the following theorem.

Theorem 1: The  necessary and sufficient condition for hydrostatic equilibrium of any static\footnote[2]{the notion 
`any static' instead of `any regular' is used here in the
general sense, since the dynamical stability of singular solutions (which may also satisfy the compatibility criterion)
does not correspond to any solution.} 
and spherical configuration is that for an assigned value of the ratio of central pressure to
central energy-density, the compactness ratio $u(\equiv M/R)$ of the said configuration should not exceed
the compactness ratio $u_h$ of the corresponding sphere of homogeneous density distribution.

\section{Necessary and sufficient condition for hydrostatic equilibrium and dynamical stability
 of regular configuration}

The {\em absolute} values are
obtained by using a `compressible' sphere of homogeneous energy-density
(Negi 2004a), such that the following relation holds good for a constant $\Gamma_1$

\[ \frac {\Gamma_1P}{P + E} = \frac {dP}{dE}. \] 

{\noindent}
And the adiabatic speed of sound, $v = \sqrt (dP/dE)$, becomes finite inside this configuration
for a finite (constant) $\Gamma_1$. In order to satisfy the extreme case of causality condition
$v = c = 1$ at the centre of this sphere, we obtain $(P_0/E_0) \cong 0.6271$, which correspond to 
a $u$ value
$\cong 0.3406$, and the (critical) constant $\Gamma_1 = (\Gamma_1)_0 \cong 2.5946$ respectively, for
the dynamically stable configuration. This value of $u (\cong 0.3406)$ represents an absolute
upper bound, consistent with causality and dynamical stability, since it follows from the compatibility criterion
that for this maximum value of $u$, the corresponding value of $(P_0/E_0)$ of any regular configuration
can not be less than 0.6271. Now, this result may be generalized for the sequences, composed of NS models such that
every member of this sequence satisfies $(dP/dE)_0 = 1$ (here and elsewhere in the paper, the subscript `0' represent the value of
the corresponding quantity at the centre), in the following manner that the maximum stable value of $u$
(corresponding to
the case of first maxima among masses in the $M-R$ relation) and the corresponding central value
of local $(\Gamma_1)[= (\Gamma_1)_0]$ of such sequences must satisfy the inequalities $u_{\rm max} \leq 0.3406$,
and $(\Gamma_1)_0 \leq 2.5946$ respectively, in order to ensure the necessary and
sufficient condition for dynamical stability of a mass. Since these absolute values are obtained by
using the `trial function', $\xi = re^{\nu/4}$, which is able to provide the necessary and sufficient condition 
for dynamical stability \footnote[3]{although, the variational method gives only a sufficient
condition for the dynamical stability of a mass, since the results depend somewhat on the choice of a particular
`trial function'. However, we have shown that the choice of the particular trial function, $\xi = re^{\nu/4}$,
is capable of providing the most rigorous results among the various trial functions (Negi \& Durgapal 1999b; Negi 2004a), and because of 
this reason it could provide the necessary and sufficient condition for dynamical stability. It would not
be out of place here to point out that the variational method gives precise results only for the configurations
which correspond to a smooth variation of density from centre to the surface (see, e.g. Bardeen et. al 1966), and for this reason
we did not consider this method for analyzing the stability of core-envelope models considered here.}
in the variational method (Chandrasekhar 1964a, b).

The equilibrium sequences of the type mentioned here, in fact, are widely discussed in the literature,
but not on the basis of compatibility criterion.
The core-envelope models presented by Negi \& Durgapal (2000) also represent an equilibrium
sequence of this type. We re-constructed this sequence on the basis of compatibility criterion for the first time (for other
 example of such a sequence, consistent with the compatibility criterion, 
see, e.g. Negi 2005), and it is seen that the maximum stable value of compactness ratio and the corresponding central value
of local $(\Gamma_1)[= (\Gamma_1)_0]$ of such sequences are found fully consistent with those of the 
values obtained mentioned above (section 4; see also Negi 2005). It follows, therefore, that the $M-R$ relation
(or the mass-central density relation) provides the necessary and sufficient condition for dynamical stability
of a mass only when the compatibility criterion for the equilibrium configuration is also satisfied. Or, in other words, the compatibility criterion
is able to provide the necessary and sufficient condition for hydrostatic
equilibrium of regular configurations.

In order to verify the last claim, irrespective of the particular type of core-envelope models considered 
in the present study, we would further consider (section 5) the dynamical stability of pure polytropic configurations ($P = K\rho^{\Gamma_1}$)
on the basis of variational method for the choice of trial function $\xi = re^{\nu/4}$, as well as the mass-central density
relation for some assigned values of constant $\Gamma_1$, since the polytropic configurations appropriately
satisfy the `compatibility criterion' (Negi \& Durgapal 2001).

The results which would follow from the study of sections 4 and 5 respectively, may be summarized
as theorem 2 and its subsequent corollaries in the following form

Theorem 2: The mass-radius (or, mass-central density) relation for an equilibrium sequence of
regular configurations provides the necessary and sufficient condition for dynamical stability only
when the equilibrium sequence it-self satisfies the necessary and sufficient condition of hydrostatic 
equilibrium (theorem 1).

Corollary 1 to theorem 2: If an equilibrium sequence, composed of neutron star models 
in such a manner that every member of this sequence
satisfies the extreme causality condition $v = c = 1$ at the centre, then the maximum value of compactness
ratio (corresponding to
the case of first maxima among masses in the $M-R$ relation) and the corresponding central value
of local adiabatic index $(\Gamma_1)_0$ are constrained by the inequalities, $u_{\rm max} \leq 0.3406$
and $(\Gamma_1)_0 \leq 2.5946$ respectively.

Corollary 2 to theorem 2: For regular configurations, corresponding to enough smooth density variations
such that the variational method could be used, the variational method could provide the necessary and sufficient condition 
of dynamical stability just for the choice of a particular trial function $\xi = re^{\nu/4}$.


\section{Hydrostatic equilibrium and dynamical stability of core-envelope models}
The metric for spherically symmetric and static configurations can be written
in the following form
\begin{equation}
ds^2 =  e^{\nu} dt^2 - e^{\lambda} dr^2 - r^2 d\theta^2 - r^2 $ sin$^2 \theta d\phi^2 , 
\end{equation}
where $\nu$ and $\lambda$ are functions of $r$ alone. Recalling that we are using `geometrized units', the  Oppenheimer-Volkoff 
(O-V) equations (Oppenheimer \& Volkoff 1939), resulting  from  Einstein's field  equations, for 
systems with isotropic pressure $P$  and  energy-density  $E$  can  be 
written as
\begin{eqnarray}
P' & = & - (P + E)[4 \pi P r^3 + m]/r(r - 2m) \\                       
\nu'/2 & = & - P'/(P + E)  \\                                            
m'(r) & = & 4\pi E r^2 \,;
\end{eqnarray}
where $ m(r) $ is the mass contained within the  radius  $r$,  and  the 
prime denotes radial derivative.

The core-envelope model (Negi \& Durgapal 2000) consists of a core  with  most  stiff 
EOS in the region $0 \leq r \leq b$, and an envelope with a polytropic  EOS 
in the region $b \leq r \leq R$, given as 

\medskip

(i) The core: $0 \leq r \leq b$

\medskip

For the models of neutron  stars  considered  here,  we  have 
chosen the core of most stiff material as
\begin{equation}
 P  =  (E - E_s)
\end{equation}
where  $E_s$  is  the  value  of  density  at  the  surface  of   the 
configuration, where pressure vanishes.

\medskip

(ii) The envelope: $b \leq r \leq R$

\medskip

The envelope of this model is given by the equation of state
\begin{equation}
P  =  K \rho^{\Gamma_1}
\end{equation}
or
                                                
$(E - \rho)  =  P/(\Gamma_1 - 1)$.

\medskip

where $K$ is a constant to be worked out by the matching of  various 
variables at the core-envelope boundary and $\rho$ and $\Gamma_1$ represent respectively, the rest-mass density and the (constant) adiabatic index as defined earlier.

\medskip

At the boundary, $r = b$, the continuity of $P(=P_b), E(=E_b)$, and 
$r(=r_b)$ require
\begin{equation}
K = P_b /{(E_b  - [P_b /(\Gamma_1 - 1)])}^{\Gamma_1}
\end{equation}
                                   
where $\Gamma_1$  is given by (see, e.g., Tooper 1965)

\medskip

$\Gamma_1  = [(P + E)/P](dP/dE)$.

\medskip

The continuity of $(dP/dE)$, at the boundary gives
\begin{equation}
\Gamma_1  = 1 + (E_b/P_b).
\end{equation}
Thus,  the  continuity  of  $P, E, \nu, \lambda,$  and $(dP/dE)$  at  the 
core-envelope boundary is ensured, for the static and spherically 
symmetric configuration.

The coupled Eqs. (15), (16), (17), are  solved  along  with 
Eqs.(18) and (19) for the boundary conditions (20) and (21)  [at 
the  core-envelope  boundary, $ r  =   b $],    and   the    boundary  
conditions, $P  = E = 0$ , $m(r = R)  =  M$, $e^{\nu} = e^{-\lambda} = (1 - 2M/R) = 
(1 - 2u)$ at the external boundary, $r  =  R$. 


For the sake of  numerical  simplification,  we  assign  the 
central density, $E_0 = 1$. It is seen that the degree  of  softness 
of the envelope is restricted by the inequality, $(P_b/E_b) \geq 0.014$. 
For the minimum value of $(P_b/E_b) \cong  0.014$,  we  obtain  various 
quantities, such as, core mass, $ M_b $, core radius, $r_b$ ,  density  at 
the  core-envelope  boundary, $E_b$ ,  total  mass, $M$,   and   the 
corresponding radius, $R$, of the  configuration  in  dimensionless 
form. Some of these quantities are shown in Table 1  for  various  assigned  values  of  the 
central pressure to density ratio, $(P_0/E_0)$. 

To determine the stability of the models given in  Table  1, 
we need to draw the mass-radius diagram for the  structures.  For 
this purpose, we have normalized the boundary density, $E_b = 2 \times 10^{14}$ g\, cm$^{-3}$, 
and obtained the mass-radius diagram as shown in Fig. 1 of Negi and Durgapal (2000)
[Notice that the value of $E_b$ chosen in this way (and hence also 
the mass and the radius obtained in conventional units  as  shown 
in Fig. 1) is purely arbitrary. These values have nothing  to  do 
with the actual maximum mass and the corresponding radius of  the 
stable neutron star obtained in the present paper.]. The maximum  stable  value 
of $u$ of the whole configuration is obtained  as  0.3574. For this maximum value  of $u$, 
the binding energy per baryon, $\alpha_r[\equiv (M_r - M)/M_r$, where $M_r$ is  the 
rest-mass (Zeldovich \& Novikov 1978) of the configuration]  also 
approaches maximum ($\cong 0.2441$) as shown in Table 1. Although, the corresponding $P_0/E_0 (\simeq 0.704)$ value
(or $(\Gamma_1)_0 \simeq 2.4204$ value) is consistent with the corresponding absolute value ($P_0/E_0$ is larger
than 0.6271 or $(\Gamma_1)_0$ is less than 2.5946), the configuration is not consistent with the corollary 1
 of theorem 2, since the maximum value of $u \simeq 0.3574$ is inconsistent with the absolute upper bound on $u \cong 0.3406$.
It follows, therefore, that the $M-R$ relation does not provide the necessary and sufficient condition
for dynamical stability of equilibrium masses, since these models are not consistent with the compatibility 
criterion, which is evident from Table 1. As the first column of Table 1 corresponds to compactness ratio, $u_h$, of homogeneous
density sphere as calculated from Eqs. (1) and (2) for various assigned values of
$\sigma$ shown in Table 1. Column seventh of this table represents the compactness
ratio $u$ of the whole configuration for the same values of $\sigma$. Comparing column
one and seventh, we find that for each assigned value of $\sigma, u > u_h$, meaning
thereby that the model is inconsistent with the compatibility 
criterion.

To make the model consistent with the structure of general relativity, we re-investigate
this model based upon the compatibility criterion (section 2) by solving the coupled
Eqs. (15), (16), and (17) together with Eqs. (18) and (19) for the boundary conditions
(20) and (21) respectively. For numerical simplicity, we assign the central energy-density
of the configuration, $E_0 = 1$. The order of numerical precision is set precisely
following the specific nature of EOSs for the core and envelope regions respectively. The ratio, $(P_b/E_b)$, at the core-envelope boundary
 is so adjusted that for each and every (possible) assigned value of $\sigma$, the compactness
 ratio, $u$, of the whole configuration always turns out to be less than or equal to the
 compactness ratio, $u_h$, of the homogeneous density sphere for same values of $\sigma$.
The results obtained in this regard are shown in Table 2. It is seen that to meet the requirement
set up by compatibility criterion, the minimum value of $(P_b/E_b)$ reaches about
$2.9201 \times 10^{-1}$.

To investigate the stability of the models which are now compatible with the structure of general relativity
and causality (Table 2), we draw the mass-radius diagram for the models by normalizing the 
boundary density, $E_b = 2 \times 10^{14} {\rm g\, cm}^{-3}$, as shown in Fig. 1
[notice that the use of normalizing density, $E_b$, as mentioned in the previous case also, is
purely arbitrary and its purpose is only to determine the maximum value of $u$ upto
which the structures remain pulsationally stable]. The {\em first} maxima in mass,
among the equilibrium sequences of masses, is reached when the ratio of central pressure
to central energy-density, $\sigma [\equiv (P_0/E_0)]$, approaches to a value about
0.6285. The binding-energy per baryon, $ \alpha_r[\equiv (M_r - M)/M_r$ ] also approaches to its {\em first} maxima for this maximum stable value of mass. 
The corresponding maximum stable value of compactness ratio, $u$, is obtained as 0.3389 (Table 2).
Thus, the structure remains pulsationally stable upto a $u$ value as large as 0.3389, that is, 
$u \leq 0.3389$ [notice that both, the upper bound on maximum value of $u < 0.3406$, and $(\Gamma_1)_0 < 2.5946$ are fully
consistent with the corresponding absolute upper 
bounds]. Thus, the corollary 1 of theorem 2 is fully satisfied for this case. It follows, therefore, that
the $M-R$ relation provides the necessary and sufficient condition
for dynamical stability of equilibrium masses, since these models are fully consistent with the compatibility 
criterion. This is evident from the comparison of column one and column seventh of Table 2 which indicates
that for each value of $(P_0/E_0)$, the compactness ratio of the whole configuration, $u$, is always less
than $u_h$, the compactness ratio of the corresponding homogeneous density sphere. Obviously, the upper
bound on $u \leq 0.3389$ obtained here by using the compatibility criterion, turns out to be much
stronger than the upper bounds on this parameter obtained by Lindblom (1984) and Haensel et al (1999).

\section{Dynamical stability of polytropic configurations}

The dynamical stability of polytropic configurations ($P = K\rho^{\Gamma_1}$) was first investigated
by Tooper (1965) for polytropic index $3 \leq n \leq 1 (\Gamma_1 = 1 + 1/n)$ by using the variational method which states that a {\it sufficient} condition for the dynamical stability of a mass is that
the right-hand side of the following equation 

\begin{eqnarray}
\lefteqn{\omega^2 \int_{0}^{R} e^{(3\lambda - \nu) /2} (P + E) r^2 \xi^2 dr = }  \nonumber  \\
  & & 4 \int_{0}^{R} e^{(\lambda + \nu)/2} r P'\xi^2 dr  \nonumber  \\
  & & + \int_{0}^{R} e^{(\lambda + 3 \nu)/2} [\gamma P/r^2] {(r^2 e^{-\nu/2} \xi)'}^2 dr  \nonumber  \\
  & & - \int_{0}^{R} e^{(\lambda + \nu /2)} [P'^2/(P + E)] r^2 \xi^2 dr  \nonumber  \\
  & & + 8\pi \, \int_{0}^{R} e^{(3\lambda + \nu) /2} P (P + E) r^2 \xi^2 dr.   
\end{eqnarray}

vanishes for some chosen ``trial function'' $\xi$ which
satisfies the boundary conditions
\begin{eqnarray}
\xi & = & 0 \hspace{.2in} {\rm at} \hspace{.2in} r = 0,
\end{eqnarray}
and
\begin{eqnarray}
\delta P & = & - \xi P' - \gamma P e^{\nu/2} [(r^2 e^{-\nu/2} \xi)'/r^2]  \nonumber  \\
         & = & 0 \hspace{.2in} {\rm at} \hspace{.2in} r = R,
\end{eqnarray}
where $\omega$ is the angular frequency of pulsation, $ R $ is the size of the configuration, 
and $\delta P$ is the `Lagrangian 
displacement  in  pressure'. The   prime   denotes   radial 
derivative, and the quantity $\gamma = [(P + E)/P](dP/dE) = \Gamma_1$ (constant) for the polytropic configurations
considered here. Tooper (1965) used the trial function of the form $\xi = b_1 r(1 + a_1 r^2  + a_2 r^4  + a_3 r^6  + ...)e^{\nu/2}$, 
where $a_1, a_2, a_3,$ ... are adjustable constants, in Eq.(22) and showed that for $3 < n \leq 1$, the
first maxima among the masses in the mass-central density (or, mass-radius) relation approaches at the 
same value of central pressure to central rest-density ratio ($P_0/\rho_0$) where the squared frequency of pulsation, $\omega^2$, also
becomes zero.

In order to verify these results in view of the discussion of section 3, we choose the trial function $\xi = re^{\nu/4}$
and employ a fourth-order Runge-Kutta method to solve Eq.(22). The results of this iteration are presented in
Tables 3-4 for the polytropic index $n = 1$ and 1.5 respectively. It is seen that the first maxima among masses
in the mass-centre density relation is reached for the same value of $(P_0/\rho_0)$ where $\omega^2$ also
approaches zero. This finding is in perfect agreement with those of the Tooper (1965) and together with
theorem 2 verifies further
that just the choice of the trial function $\xi = re^{\nu/4}$ is capable of providing the necessary and sufficient
condition for dynamical stability of masses.

\section{Results and conclusions}

We have re-investigated the core-envelope model with  stiffest  equation  of  state 
[speed of sound equal to  that  of  light]  in  the  core  and  a 
polytropic  equation  with  constant   adiabatic   index  $\Gamma_1 = [$d$lnP/$d$ln\rho]$ in the envelope,
based upon the criterion obtained by Negi and Durgapal (2001).
We find that the condition of hydrostatic equilibrium is assured only when the minimum
ratio of $(P_b/E_b)$ at the core-envelope boundary reaches about $2.9201 \times 10^{-1}$ [that is, 
when the value of the adiabatic index, $\Gamma_1$ at the core-envelope boundary reaches around 4.4246
as compared to the previous case of $\Gamma_1 \cong 72.4286$].
Under this condition, the pressure, density, both of the metric parameters  including  their 
first derivatives, and the speed of sound are  continuous  at  the 
core-envelope boundary  and  at  the  surface. The mass-radius diagram indicates
that the configuration remains dynamically stable upto a $u$ value as large as
0.3389. The corresponding central value of the local $(\Gamma_1)_0$ is obtained as 2.5911. These values are fully consistent 
with those of the {\em absolute} values,
$u_{max} \cong 0.3406$, and $(\Gamma_1)_{0,{\rm max}} \cong 2.5946$, compatible with the structure of general relativity, causality, and
dynamical stability, obtained by using a causal configuration of homogeneous energy-density (Negi 2004a).
Not just for the particular model considered in the present study, we have also found that the maximum
value of $u$ and the corresponding value of $(\Gamma_1)_0$ upto which the configurations remain pulsationally stable can not exceed the values of 
$u \simeq 0.34$ and $(\Gamma_1)_0 \simeq 2.5128$ respectively,
if the `compatibility criterion' is followed and the envelope of the present model is replaced by the
polytropic EOS $ (4/3) \leq $d$lnP/$d$ln\rho \leq 2 $ (Negi 2005). 

In addition to the study of two-density
models considered in the present study (sec.4), the dynamical stability of the polytropic configurations (sec.5)
 explicitly shows that the compatibility criterion (Negi \& Durgapal 2001)
alone is capable of providing a necessary and sufficient condition for any regular
configuration to be consistent with the structure of general relativity [however, the study of $M-R$ relation of such
sequences (corresponding to two-density models with $v = c = 1$ at the centre), consistent with the definition of 
actual mass (Negi 2004 b; Negi 2006) and the `compatibility
criterion' (Negi \& Durgapal 2001) in the near future will finally settle down this issue].

 The two-density  structures   are   dynamically   stable    and 
gravitationally bound even for the value of compactness ratio, 
$u \geq  (1/3)$,  thus  giving  a  suitable  model  for  studying   the 
Ultra-compact Objects [UCOs] discussed in the literature (see, e. g., Negi \& Durgapal 1999a, b; 2000;
and references therein). The present type of studies may also
find application to test various models of NSs based upon EOSs of dense nuclear matter, and the models of relativistic
stellar objects like - star clusters. 

\acknowledgments The author acknowledges the Aryabhatta Research Institute of Observational Sciences (ARIES), Nainital for providing 
library and computer-centre facilities.

\newpage

Table 1: Properties of the causal core-envelope models, with a core given by the most
              stiff EOS, $ (dP/dE) = 1 $, and the envelope is characterized by
              the polytropic EOS, (d$ lnP/$d$ ln \rho) = \Gamma_1 $, such that, all
              the parameters, $P, E, \nu, \lambda,$ and the speed of sound, ${(dP/dE)}^{1/2} $,
              are continuous at the core-envelope boundary, $r_b$, and the models satisfy the necessary (but not sufficient)
              condition for hydrostatic equilibrium. The maximum value of $ u[ \equiv (M/R) \cong 0.3574] $
              for the structure is obtained [Fig. 1 of Negi \& Durgapal (2000)], when the minimum value of the ratio of pressure to density at the
              core-envelope boundary, $ (P_b/E_b) $, reaches about 0.014. The maximum value of the binding-energy
              per baryon, $ \alpha_r[\equiv (M_r - M)/M_r, $ where $M_r$ is the rest mass of the configuration]
              $ \cong 0.2441, $ also occurs for the maximum stable value of $u$. The subscript `0' and `b' represent,
               the values of respective quantities at the centre, and at the core-envelope boundary. $z_R$
               stands for the surface redshift. The calculations are performed for an assigned value
               of the central energy-density, $E_0 = 1$. Various values shown in the table are round off at the
               fourth decimal place. The slanted values represent the limiting case
               upto which the structure remains dynamically stable. However, the model
               do not satisfy the necessary and sufficient
              condition for hydrostatic equilibrium, since 
               for each assigned value of $(P_0/E_0)$, the compactness ratio, $u$,
               of the whole configuration always corresponds to a value {\em larger} than that of
               the compactness ratio, $u_h$, of the homogeneous density distribution (that is $u > u_h$). Furthermore,
               the $M-R$ relation does not provide the
               necessary and sufficient condition for dynamical stability of equilibrium configurations,
               since the maximum stable value of $u \simeq 0.3574$ exceeds the limiting value ($u\cong 0.3406$)
               for the corresponding centre value of $(\Gamma_1)_0 \simeq 2.4204$ (which is, however,
               consistent with the corresponding absolute upper bound $\leq 2.5946$).

\begin{table*}
\begin{center}

\hspace{1.0cm}

\begin{tabular}{cccccccccc}


$u_h$ & ${P_0 / E_0}$ & $(\Gamma_1)_0$ & $r_b$    & $E_b$    & $R$   & $u$  & $\alpha_r$
 & $z_R$   & $z_0$  \\


0.1527 & 0.1110  & 10.009 & 0.1886 & 0.9017  & 0.2012  & 0.1558 & 0.1001 & 0.2052
& 0.3473  \\

0.1901 & 0.1562 & 7.4020 & 0.2181 & 0.8559  & 0.2282 & 0.1946 & 0.1290 & 0.2795 &0.4977 \\

0.2195 & 0.2012 & 5.9702 & 0.2410 & 0.8101  & 0.2489 & 0.2243 & 0.1512 & 0.3467 &0.6515 \\

0.2478 & 0.2564 & 4.9002 & 0.2639 & 0.7542  & 0.2712 & 0.2550 & 0.1751 & 0.4287 &0.8573 \\

0.2696 & 0.3100 & 4.2258 & 0.2832 & 0.6998  & 0.2887 & 0.2763 & 0.1917 & 0.4949 &1.0598 \\

0.2975 & 0.4000 & 3.5000 & 0.3127 & 0.6086  & 0.3176 & 0.3058 & 0.2152 & 0.0648 &1.4513 \\

0.3213 & 0.5070 & 2.9724 & 0.3477 & 0.5001 & 0.3522  & 0.3304 & 0.2334 & 0.7168 &2.0020 \\

0.3315 & 0.5661 & 2.7665 & 0.3690 & 0.4401  & 0.3733 & 0.3405 & 0.2395 & 0.7703 &2.3633 \\

0.3369 & 0.6010 & 2.6639 & 0.3829 & 0.4047  & 0.3870 & 0.3455 & 0.2419 & 0.7988 &2.6033 \\

0.3424 & 0.6410 & 2.5601 & 0.4003 & 0.3642  & 0.4043 & 0.3504 & 0.2434 & 0.8282 &2.9084 \\

0.3501 & {\sl 0.7040} & {\sl 2.4204} & {\sl 0.4329} & {\sl 0.3002}  & {\sl 0.4376} & {\sl 0.3574} & {\sl 0.2441} & {\sl 0.8724} & {\sl 3.4925 } \\

0.3504 & 0.7070 & 2.4144 & 0.4347 & 0.2971  & 0.4395 & 0.3577 & 0.2441 & 0.8744 &3.5242 \\

0.3513 & 0.7151 & 2.3984 & 0.4396 & 0.2890  & 0.4443 & 0.3583 & 0.2436 & 0.8783 &3.6085 \\

0.3523 & 0.7238 & 2.3816 & 0.4450 & 0.2801  & 0.4497 & 0.3590 & 0.2430 & 0.8833 &3.7055 \\


\end{tabular}

\end{center}
\end{table*}

\newpage

Table 2: Properties of the causal core-envelope models, as discussed in the present paper,
      with a core given by the most
              stiff EOS, $ (dP/dE) = 1 $, and the envelope is characterized by
              the polytropic EOS, (d$ lnP/$d$ ln \rho) = \Gamma_1 $, such that, all
              the parameters, $P, E, \nu, \lambda,$ and the speed of sound, ${(dP/dE)}^{1/2} $,
              are continuous at the core-envelope boundary, $r_b$. The maximum value of $ u[ \equiv (M/R) \cong 0.3389] $
              for the structure is obtained (Fig. 1), when the minimum value of the ratio of pressure to density at the
              core-envelope boundary, $ (P_b/E_b) $, reaches about $2.9201 \times 10^{-1}$. The {\em first} maxima among the 
              values of the binding-energy
              per baryon, $ \alpha_r[\equiv (M_r - M)/M_r, $ where $M_r$ is the rest mass of the configuration]
              also occurs for the maximum stable value of $u$.  The calculations are performed for an assigned value
               of the central energy-density, $E_0 = 1$. Except $(P_0/E_0)$
               and $r_b$, all other values are round off at the fourth decimal place. The subscript `0' and `b' represent,
               the values of respective quantities at the centre, and at the core-envelope boundary. $z_R$
               stands for the surface redshift. The slanted values represent the limiting case
               upto which the structure remains dynamically stable. The model
               is fully compatible with the structure of general relativity, as
               it is seen that for each assigned value of $(P_0/E_0)$, the compactness ratio, $u$,
               of the whole configuration always corresponds to a value {\em less than or equal to} that of
               the compactness ratio, $u_h$, of the homogeneous density distribution (that is $u \leq u_h$). 
               The maximum stable value of $u\simeq 0.3389 (< 0.3406)$ and the corresponding central value
               of $(\Gamma_1)_0 \simeq 2.5911 (< 2.5946)$ indicate that the $M-R$ relation provides the
               necessary and sufficient condition for dynamical stability of equilibrium configurations.

\newpage 

\hspace{1.0cm}
\begin{table*}
\begin{center}
\begin{tabular}{cccccccccc}


$u_h$ & ${P_0 / E_0}$ & $(\Gamma_1)_0$ & $r_b$    & $E_b$    & $R$   & $u$  & $\alpha_r$
 & $z_R$   & $z_0$  \\
$(10^0)$& $(10^0)$ &  $(10^0)$ & $(10^{-3})$& $(10^0) $ & $(10^0)$ & $(10^0)$ & $(10^0)$ & $(10^0)$ & $(10^0)$ \\

0.2628 & 0.29202  & 4.4244 & 2.68 & 0.9999  & 0.3117 & 0.2628 & 0.1698 & 0.4520 & 0.9615 \\

0.2652 & 0.29800 & 4.3557 & 41.25 & 0.9915  & 0.3132  & 0.2649 & 0.1714 & 0.4584 & 0.9831 \\

0.2846 & 0.35452 & 3.8207 & 128.53 & 0.9117 & 0.3275  & 0.2837 & 0.1870 & 0.5205 & 1.2026 \\

0.2976 & 0.39993 & 3.5005 & 166.04 & 0.8476  & 0.3400  & 0.2965 & 0.1984 & 0.5676 & 1.3943 \\

0.3072 & 0.43923 & 3.2767 & 191.90 & 0.7920  & 0.3509  & 0.3063 & 0.2069 & 0.6066 & 1.5738 \\

0.3112 & 0.45699 & 3.1882 & 202.42 & 0.7670  & 0.3564  & 0.3102 & 0.2105 & 0.6232 & 1.6588 \\

0.3188 & 0.49397 & 3.0244 & 222.83 & 0.7147  & 0.3678  & 0.3178 & 0.2170 & 0.6568 & 1.8467 \\

0.3261 & 0.53331 & 2.8751 & 243.11 & 0.6592  & 0.3801  & 0.3252 & 0.2224 & 0.6913 & 2.0657 \\

0.3274 & 0.54099 & 2.8485 & 246.97 & 0.6483  & 0.3830  & 0.3265 & 0.2235 & 0.6974 & 2.1101\\

0.3306 & 0.55999 & 2.7857 & 256.44 & 0.6215  & 0.3894  & 0.3296 & 0.2256 & 0.7130 & 2.2255\\

0.3355 & 0.59146 & 2.6907 & 272.04 & 0.5770  & 0.4020  & 0.3341 & 0.2289 & 0.7361 & 2.4265\\

0.3394 & 0.61902 & 2.6154 & 285.84 & 0.5381  & 0.4128  & 0.3380 & 0.2307 & 0.7568 & 2.6215\\

0.3407 & {\sl 0.62850} & {\sl 2.5911} & {\sl 290.65} & {\sl 0.5247}  & {\sl 0.4176} & {\sl 0.3389} & {\sl 0.2314} & {\sl 0.7618} & {\sl 2.6888} \\

0.3409 & 0.62936 & 2.5889 & 291.09 & 0.5235  & 0.4171 & 0.3394 & 0.2311  & 0.7642 & 2.6989 \\

0.3411 & 0.63148 & 2.5836 & 292.18 & 0.5205 & 0.4182  & 0.3396 & 0.2313  & 0.7654 & 2.7146\\

0.3438 & 0.65148 & 2.5350 & 302.55 & 0.4922 & 0.4276  & 0.3419 & 0.2322  & 0.7783 & 2.8710\\

0.3497 & 0.69998 & 2.4286 & 329.29 & 0.4238 & 0.4533  & 0.3467 & 0.2328  & 0.8061 & 3.2991\\

0.3553 & 0.75294 & 2.3281 & 362.83 & 0.3490 & 0.4888  & 0.3505 & 0.2306  & 0.8290 & 3.8720\\


\end{tabular}

\end{center}
\end{table*}

\newpage

Table 3: Properties of the models characterized by
              the pure polytropic EOS, (d$ lnP/$d$ ln \rho) = \Gamma_1 $, for $\Gamma_1 = 5/3 (n = 1.5)$. 
              The 
              stability of the models is judged by the variational method for the choice of the trial
              function $\xi = re^{\nu/4}$ in Eq.(22), as well as the mass-central density relation
               [equivalent to dimensionless mass ($M/M^{\ast})$ vs. $(P_0/E_0)$ or $(P_0/\rho_0)$ ratio; where $M^{\ast} = 
               (n + 1)^{3/2}(P_0/\rho_0)^{n/2}/(4\pi\rho_0)^{1/2}$]. It is apparently seen that the configurations
               become dynamically unstable beyond the maximum mass where the squared angular
               frequency of pulsation, $\omega^2$, also approaches zero. Thus, it follows that the mass-central density relation 
               provides the necessary and sufficient condition for dynamical stability of equilibrium configurations. The maximum 
              value of the binding-energy
              per baryon, $ \alpha_r[\equiv (M_r - M)/M_r, $ where $M_r$ is the rest mass of the configuration]
              also occurs for the maximum value of mass. All values are round off at the fourth decimal
               place. The subscript `0' represents,
               the values of respective quantities at the centre. The slanted values represent the limiting case
               upto which the structure remains dynamically stable. The model
               is fully compatible with the structure of general relativity, as
               it is seen that for each assigned value of $(P_0/E_0)$, the compactness ratio, $u$,
               of the whole configuration always corresponds to a value {\em less than} that of
               the compactness ratio, $u_h$, of the homogeneous density distribution (that is $u < u_h$). 

\newpage 

\hspace{1.0cm}
\begin{table*}
\begin{center}
\begin{tabular}{ccccccc}


${P_0 / E_0}$ & $P_0/\rho_0$ & $M/M^\ast$ & $\alpha_r$ & $\omega^2/E_0$  & $u$ & $u_h$ 
   \\


0.0465 & 0.0500 & 0.1907 & 0.0277 & 1.1085  & 0.0698 & 0.0783\\

0.0653 & 0.0724 & 0.2153 & 0.0343 & 0.8047  & 0.0904 & 0.1032 \\

0.0805 & 0.0916 & 0.2272 & 0.0381 & 0.5734 & 0.1046  & 0.1213 \\

0.1026 & 0.1213 & 0.2359 & 0.0414 & 0.2669  & 0.1221 & 0.1446 \\

0.1158 & 0.1402 & 0.2378 & 0.0422 & 0.0979  & 0.1308 & 0.1571 \\

{\sl 0.1239} & {\sl 0.1522} & {\sl 0.2380} & {\sl 0.0424} & {\sl 0.0000}  & {\sl 0.1356} & 0.1643 \\

0.1300 & 0.1615 & 0.2377 & 0.0422 & -0.0715  & 0.1391 & 0.1695 \\

0.1523 & 0.1974 & 0.2346 & 0.0405 & -0.3130  & 0.1501 & 0.1872 \\

0.1798 & 0.2462 & 0.2273 & 0.0357 & -0.5694  & 0.1604 & 0.2061 \\

0.2696 & 0.4526 & 0.1924 & 0.0021 & -1.0936  & 0.1750 & 0.2537 \\

0.3279 & 0.6453 & 0.1684 & -0.0324 & -1.1804  & 0.1716 & 0.2760 \\

0.3705 & 0.8340 & 0.1523 & -0.0624 & -1.1146  & 0.1635 & 0.2894 \\


\end{tabular}

\end{center}
\end{table*}

\newpage 

Table 4: Properties of the models characterized by
              the pure polytropic EOS, (d$ lnP/$d$ ln \rho) = \Gamma_1 $, for $\Gamma_1 = 2 (n = 1) $. The 
              stability of the models is judged by the variational method for the choice of the trial
              function $\xi = re^{\nu/4}$ in Eq.(22), as well as the mass-central density relation
               [equivalent to dimensionless mass ($M/M^{\ast})$ vs. $(P_0/E_0)$ or $(P_0/\rho_0)$ ratio; where $M^{\ast} = 
               (n + 1)^{3/2}(P_0/\rho_0)^{n/2}/(4\pi\rho_0)^{1/2}$]. It is apparently seen that the configurations
               become dynamically unstable beyond the maximum mass where the squared angular
               frequency of pulsation, $\omega^2$, also approaches zero. Thus, it follows that the mass-central density relation 
               provides the necessary and sufficient condition for dynamical stability of equilibrium configurations.
             The maximum 
              value of the binding-energy
              per baryon, $ \alpha_r[\equiv (M_r - M)/M_r, $ where $M_r$ is the rest mass of the configuration]
              also occurs for the maximum value of mass. All values are round off at the fourth decimal
               place. The subscript `0' represents,
               the values of respective quantities at the centre. The slanted values represent the limiting case
               upto which the structure remains dynamically stable. The model
               is fully compatible with the structure of general relativity, as
               it is seen that for each assigned value of $(P_0/E_0)$, the compactness ratio, $u$,
               of the whole configuration always corresponds to a value {\em less than} that of
               the compactness ratio, $u_h$, of the homogeneous density distribution (that is $u < u_h$).

\newpage

\hspace{1.0cm}
\begin{table*}
\begin{center}
\begin{tabular}{ccccccc}


${P_0 / E_0}$ & $P_0/\rho_0$ & $M/M^\ast$ & $\alpha_r$ & $\omega^2/E_0$    & $u$ & $u_h$
   \\


0.1114 & 0.1254 & 0.1752 & 0.0699 & 2.2225  & 0.1452 & 0.1530 \\

0.1252 & 0.1431 & 0.1830 & 0.0744 & 1.9497  & 0.1559 & 0.1655 \\

0.1579 & 0.1875 & 0.1958 & 0.0824 & 1.3440 & 0.1773  & 0.1913 \\

0.2203 & 0.2826 & 0.2055 & 0.0895 & 0.3512  & 0.2077 & 0.2301 \\

0.2309 & 0.3002 & 0.2059 & 0.0898 & 0.2030  & 0.2115  & 0.2356 \\

{\sl 0.2460} & {\sl 0.3263} & {\sl 0.2060} & {\sl 0.0899} & {\sl 0.0000}  & {\sl 0.2167} & 0.2430  \\

0.2548 & 0.3420 & 0.2058 & 0.0897 & -0.1126  & 0.2194 & 0.2471 \\

0.2800 & 0.3889 & 0.2045 & 0.0885 & -0.4151  & 0.2265 & 0.2580 \\

0.3266 & 0.4850 & 0.2000 & 0.0835 & -0.8980  & 0.2365 & 0.2755 \\

0.3948 & 0.6523 & 0.1902 & 0.0704 & -1.4356  & 0.2458 & 0.2961 \\

0.4430 & 0.7952 & 0.1820 & 0.0576 & -1.7007  & 0.2486 & 0.3080 \\

0.4917 & 0.9674 & 0.1733 & 0.0418 & -1.8762  & 0.2490 & 0.3184 \\


\end{tabular}

\end{center}
\end{table*}

\newpage
\begin{figure}
   \centering

      \caption{Mass-Radius diagram of the model corresponding to Table 2, for an assigned value of
            $E = E_b = 2 \times 10^{14}$ g\, cm$^{-3}$ at the core-envelope
            boundary $r_b$, such that the compactness ratio, $u$, of the whole configuration
            always turns out less than or equal to the compactness ratio , $u_h$, of
            homogeneous density sphere. This requirement is fulfilled only when the  ratio of pressure 
            to density
            $(P_b/E_b)$ at $r_b$ reaches to a minimum value about $2.9201 \times 10^{-1}$. The pressure, energy-density,
            $ \nu, \lambda $, and the speed of sound, $ {(dP/dE)}^{1/2} $
            are continuous at the core-envelope boundary.
              }
         \label{FigVibStab}
   \end{figure}

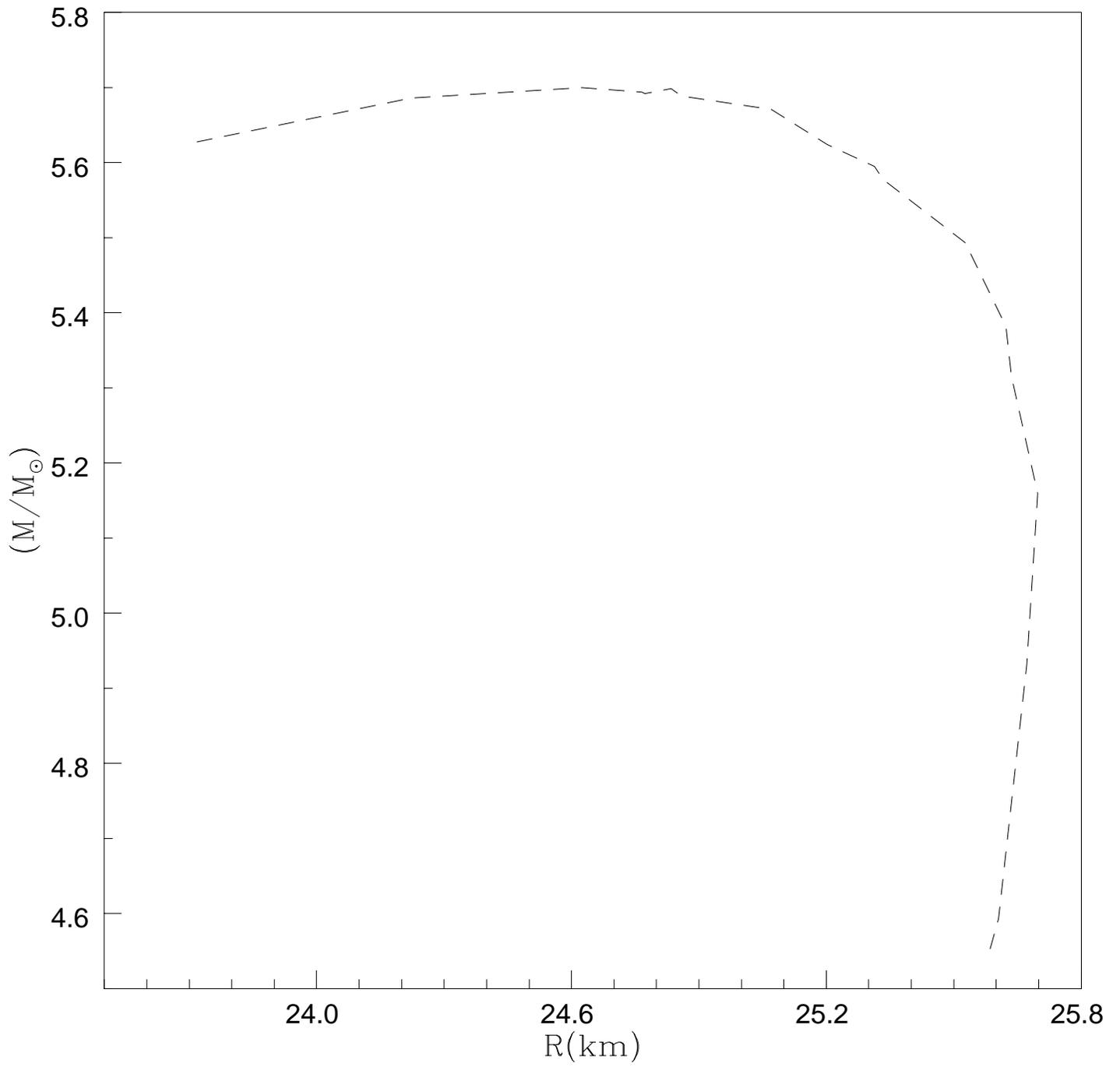
\end{document}